\documentclass[aps,epsfig,twocolumn]{revtex4}

\usepackage{graphicx}
\usepackage{dcolumn}
\usepackage{bm}

\newcommand{\bq}{\begin{equation}}
\newcommand{\eq}{\end{equation}}
\newcommand{\bqa}{\begin{eqnarray}}
\newcommand{\eqa}{\end{eqnarray}}
\newcommand{\nn}{\nonumber \\}

\begin{document}
\draft
\title{Studies of doping and temperature dependences of superfluid weight and spectral function; 
Universal scaling behavior of pseudogap}

\author{Sung-Sik Lee$^b$ and Sung-Ho Suck Salk$^{a,b,c}$}
\address{
$^a$ Korea Institute for Advanced Studies, Seoul, 130-012, Korea\\
$^b$Department of Physics, Pohang University of Science and Technology,\\
Pohang, 790-784, Korea\\
$^c$ Korean Academy of Science and Technology, Seongnam, 463-808, Korea\\
}
\date{\today}

\begin{abstract}
Using the improved U(1) and SU(2) slave-boson approaches of the t-J Hamiltonian [Phys. Rev. B {\bf 64}, 052501 (2001)] that we developed recently, 
we study the doping and temperature dependence of superfluid weight and spectral function
and discuss our finding of the universal scaling behavior of pseudogap.
It is shown that at low hole doping concentrations  $x$ and at low temperatures $T$ there exists a propensity of linear decrease in the superfluid weight $n_s/m^*$ with temperature $T$, and a tendency of doping independence in the linearly decreasing slopes of $\frac{n_s}{m^*}(x,T)$ with $T$ in qualitative agreement with the experimentally observed relation, that is, $\frac{n_s}{m^*}(x,T) = \frac{n_s}{m^*}(x,0) - \alpha T$ with $\alpha$, a constant.
It is also demonstrated that there exists the boomerang behavior, that is, both $T_c$ and $n_s/m^*$ increase with hole doping concentration $x$ in the underdoped region, reaches a saturation(maximum) at a hole doping above optimal doping and decreases beyond the saturation point in the overdoped region in agreement with $\mu$SR measurements.
Based on our improved SU(2) slave-boson approach to the t-J Hamiltonian, 
we further investigate the doping and temperature dependence of spectral function and discuss our theoretical finding of a scaling behavior of pseudogap.
In addition we discuss the cause of hump and quasi-particle peak in the observed spectral functions of high $T_c$ cuprates. 
It is demonstrated that the sharpening of the observed quasi-particle peak below $T_c$ is attributed to the bose condensation of holon pair in agreement with observations. 
From the predicted ratios of pseudogap $\Delta_0$ to both the superconducting temperature $T_c$ and the pseudogap temperature $T^*$
 i.e., $\frac{2 \Delta_0}{k_B T_c}$ and $\frac{2 \Delta_0}{k_B T^*}$ respectively as a function of hole doping concentration $x$, 
we find that there exists a universal scaling behavior (sample independence) of these ratios with doping, 
by showing a nonlinearly decreasing behavior of the former, that is, $\frac{2 \Delta_0}{k_B T_c} \sim x^{-\alpha}$ with $\alpha \sim 2$ 
and a near doping independence of the latter, that is, $\frac{2 \Delta_0}{k_B T^*} \approx 4 \sim 6$.
\end{abstract}
\maketitle

\newpage

\section{Introduction}

Presently there exist two outstanding physical problems related to superfluid weights for high $T_c$ cuprates.
One is the understanding of the so-called `boomerang' (reflex) behavior and the other, the doping independence of linear decrease of superfluid weight $\frac{n_s}{m^*}(x,T)$ vs. temperature $T$.
More than a decade ago the transverse field muon-spin-relaxation($\mu$-SR) measurements of the magnetic penetration depth $\lambda$ in high $T_c$ copper oxide superconductors revealed the `boomerang' behavior 
of the superconducting temperature $T_c$ and the superfluid weight $n_s/m^*$, 
that is, a linear increase of both $T_c$ and $n_s/m^*$ in the underdoped region, a saturation of $T_c$ near optimal doping and a decrease (reflex) of both $T_c$ and $n_s/m^*$ in the overdoped region as the hole doping concentration increases\cite{UEMURA,UEMURA_NATURE,BERNHARD}.
%%%%
Earlier Emery and Kivelson\cite{EMERY95} suggested that the linear proportionality of $T_c$ and $n_s/m^*$ in the underdoped region can be understood in terms of the phase fluctuations of Cooper pairs.
It is argued that phase fluctuations become important in the underdoped region because the superconducting charge carrier density increases linearly with hole concentration\cite{UEMURA}.
The linear increase of the superfluid weight in the underdoped region was predicted from both the slave-boson theories of the t-J Hamiltonian\cite{FUKUYAMA,KOTLIAR,NAGAOSA,UBBENS,WEN,PALEE04} and the numerical studies of projected trial wave function\cite{TRIVEDI}.
In the single holon slave-boson theories $T_c$ is given by the single holon bose condensation temperature in the underdoped region and the spin gap temperature in the overdoped region.
Similarly, the Ioffe-Larkin formula\cite{IOFFE} shows that the superfluid weight is mainly determined by the holon contribution in the underdoped region and by the spinon contribution in the overdoped region\cite{NAGAOSA}.
Thus both $T_c$ and $n_s/m^*$ are claimed to increase with hole concentration $x$ in the underdoped region and decrease in the overdoped region.
However the predicted $T_c$ increases too stiffly with $x$ in the underdoped region because $T_c$ is proportional to a large hopping
energy $t \sim 0.5 eV$\cite{HYBERTSEN} according to these  single-holon condensation theories.
In the nonlinear sigma model description of the SU(2) slave-boson theory\cite{WEN98,PALEE04} collective modes associated with order parameter fluctuations are shown to play important role in determining $T_c$ and $n_s/m^*$.
From the projected trial wavefunction studies\cite{TRIVEDI} it was predicted that the superconducting order parameter is arch-shaped as a function of hole concentration and that the upper bound of the superfluid weight increases linearly with hole concentration.
However this study is limited to the $T=0K$ and thus the relation between $T_c$ and $n_s/m^*$ has not been directly addressed.

Most recently, magnetic penetration depth($\lambda$) measurements\cite{SHENG,PANA0,PANA} showed a doping independence in the negative slopes of the superfluid weight with increasing temperature showing the relations $\frac{n_s}{m^*}(x,T) = \frac{n_s}{m^*}(x,0) - \alpha T$ in the underdoped region where $x$ is the hole concentration and $\alpha$, a temperature independent constant.
The observed ratio between the superfluid weight and the superconducting temperature $\alpha_0 = \frac{\frac{n_s}{m^*}(x,0)}{T_c(x)}$ is nearly doping independent in the underdoped region\cite{UEMURA}.
As a consequence the normalized superfluid weight displays a universal behavior, following the relation $\frac{n_s/m^*(x,T)}{n_s/m^*(x,0)} = 1 - \beta (\frac{T}{T_c})$ with $\beta = \frac{\alpha}{\alpha_0}$\cite{SHENG,PANA0,PANA}.
This universal behavior has been interpreted by the phase fluctuation model\cite{CARLSON}, or the quasiparticle model based on the Landau-Fermi liquid theory\cite{PALEE97,MILLIS,MESOT,BENFATTO}.
However, apart from these phenomenological models the microscopic explanation of this universal behavior from a microscopic theory is unresolved\cite{ORENSTEIN}.
Earlier other U(1) slave-boson theories\cite{KOTLIAR,FUKUYAMA,NAGAOSA,UBBENS,PALEE97,DHLEE} of the $t-J$ Hamiltonian predicted that at small hole doping the superfluid weight $\frac{n_s}{m^*}(x,T)$ as a function of temperature is strongly doping ($x^2$) dependent.
This is in direct conflict with the $\mu$SR measurements\cite{SHENG,PANA0,PANA}.
Lee and Wen\cite{PALEE97,WEN} proposed that the SU(2) theory which incorporates the low energy phase fluctuations of order parameters may resolve this problem.
Recently D.-H. Lee\cite{DHLEE} presented a U(1) slave-boson theory concerned with the low energy fluctuations of order parameters and the excitations of massless Dirac fermions at the d-wave nodal point.
This theory, also, showed the $x^2$ dependence of $\frac{n_s}{m^*}$ at finite temperature. 
Most recently Wang et al.\cite{WANG} also showed that the recently proposed d-density wave theory\cite{NAYAK} fails to predict the doping independence of the linearly decreasing slope of superfluid density vs. temperature.

Angle resolved photoemission spectroscopy(ARPES) measurements of high $T_c$ cuprates revealed both the temperature and doping dependence of both spectral peak intensity and pseudogap at varying pseudogap temperature $T^*$ and superconducting temperature $T_c$.
These measurements have shown a continuously increasing trend of pseudogap (leading edge gap) with decreasing temperature even below $T_c$ and the appearance of sharp peaks below $T_c$\cite{SHEN95,NORMAN}. 
Earlier, Wen and Lee\cite{WEN} reported the momentum dependence of the spectral function based on their SU(2) slave boson theory involving single-holon bose condensation.
Chubukov et al. obtained the peak-dip-hump feature of the spectral function by using their spin fermion model\cite{CHUBUKOV}.
Most recently, Muthukumar et al. studied the momentum dependence of the spectral function in the resonating-valence-bond state\cite{MUTH}. 
However there exists lack of comprehensive investigations concerning both the temperature and doping dependence of spectral functions.

In this study we plan to pay attention to these problems based on our earlier SU(2) slave-boson theory\cite{LEE} which differs from others in that coupling of the spinon pairing order $\Delta^f$ to the holon pairing order $\Delta^b$ (holon-pair channels) is essential for the formation of Cooper pairs as is well depicted in the last term in Eq.(\ref{eq:su2_holon_action}), that is,
$- \frac{J}{2} \sum_{<i,j>,\alpha,\beta} |\Delta^f_{ij}|^2 [ \Delta^{b*}_{ij;\alpha \beta } b_{\beta j} b_{\alpha i} + c.c. ]$.
Here we would like to stress that, obviously, the Cooper pair is a composite particle made of the spinon pair and holon pair in the slave-boson language.
Based on this improved treatment over a previous one\cite{GIMM} we were able to successfully reproduce the arch-shaped superconducting temperature curve in the phase diagram of high $T_c$ cuprates in agreement with observation.
Further, the application of this theory resulted in  the peak-dip-hump structure of optical conductivity again in agreement with observations\cite{LEE_OPT}.
Using this theory we present a five-fold concerted study on 
1. the temperature dependence of the superfluid weight,
2. the boomerang behavior of the superfluid weight and the superconducting temperature,
3. the origin of the hump and the sharp quasiparticle peak in observed spectral functions (ARPES)\cite{NORMAN}\cite{SHEN},
4. the doping and temperature dependence of the spectral functions, and
5. the universal (that is, sample independent) scaling behavior in the ratio of the spin gap in the spectral function to the superconducting temperature and the pseudogap temperature respectively with hole doping concentration, 
all of which are found to originate from the coupling effect of the spin (spinon pairing order) to the charge (holon pairing order) degrees of freedom.

\section{Temperature dependence of the holon and spinon pairing order parameters}

Earlier, we presented an improved U(1) and SU(2) slave-boson approach of the t-J Hamiltonian which differs from other slave-boson approaches to the t-J Hamiltonian in that coupling between the charge and spin degrees of freedom is manifested in the slave-boson representation of the Heisenberg term of $J\sum_{<i,j>} ( {\bf S}_{i} \cdot {\bf S}_{j} - \frac{1}{4}n_{i}n_{j} )$.
It is noted that the charge degree of freedom is well exhibited in the intersite charge interaction term of $\frac{1}{4}n_{i}n_{j}$\cite{LEE}.
Owing to the coupling of the holon pairs to the spinon singlet pairs as mentioned in the previous section, the Cooper pairs are formed as composite particles of the spinon pairs and holon pairs.
Holons form holon pairs via coupling to spinon pairs and the holon-pair bosons undergo bose condensation\cite{NOZIERE} at and below $T_c$.
This is the major difference of the present holon-pair boson theory as compared to other earlier theories which pay attention to singe-holon bose condensation\cite{PALEE97,DHLEE,KOTLIAR,FUKUYAMA,UBBENS,WEN,PALEE04}.
Our theory was shown to reproduce the arch shaped superconducting (holon-pair bose condensation) temperature $T_c$ as a function of doped hole concentration $x$\cite{LEE}, in agreement with the experimentally observed phase diagram.
It is, thus, of great interest to employ this theory to further check its applicability to other important areas of physics; the doping dependence of the superfluid weight and the spectral function.
Here we focus our attention to the doping and temperature dependence of the order parameters.

Since the predicted results of various physical properties are qualitatively indistinguishable between the U(1) and SU(2) slave-boson theories, below we present only the SU(2) theory for generality.
In the SU(2) slave-boson representation\cite{WEN}, the electron operator is given by $c_{\alpha}  = \frac{1}{\sqrt{2}} h^\dagger \psi_{\alpha}$ with $\alpha=1,2$, where $\psi_1=\left( \begin{array}{c} f_1 \\ f_2^\dagger \end{array} \right)$,   $\psi_2 = \left( \begin{array}{c} f_2 \\ -f_1^\dagger \end{array} \right)$ and $h = \left( \begin{array}{c}  b_{1} \\ b_{2} \end{array} \right)$ are respectively the doublets of spinon and holon annihilation operators in the SU(2) theory.
The SU(2) slave-boson representation of the t-J Hamiltonian shows\cite{LEE}
\bqa
H  & =  &  - \frac{t}{2} \sum_{<i,j>}  \Bigl[ (f_{\alpha i}^{\dagger}f_{\alpha j})(b_{1j}^{\dagger}b_{1i}-b_{2i}^{\dagger}b_{2j})  + h.c. \nn
&& + (f_{2i}f_{1j}-f_{1i}f_{2j}) (b_{1j}^{\dagger}b_{2i} + b_{1i}^{\dagger}b_{2j}) + h.c. \Bigr] \nn
 && -  \frac{J}{2} \sum_{<i,j>} ( 1 + h_{i}^\dagger h_{i} ) ( 1 + h_{j}^\dagger h_{j} ) \times \nn
 && (f_{2i}^{\dagger}f_{1j}^{\dagger}-f_{1i}^ {\dagger}f_{2j}^{\dagger})(f_{1j}f_{2i}-f_{2j} f_{1i}) -  \mu \sum_i  h_i^\dagger h_i  \nn
  && -  \sum_i  \Bigl[ i\lambda_{i}^{(1)} ( f_{1i}^{\dagger}f_{2i}^{\dagger} + b_{1i}^{\dagger}b_{2i}) + i \lambda_{i}^{(2)} ( f_{2i}f_{1i} + b_{2i}^\dagger b_{1i} ) \nn
  && + i \lambda_{i}^{(3)} ( f_{1i}^{\dagger}f_{1i} -  f_{2i} f_{2i}^{\dagger} + b_{1i}^{\dagger}b_{1i} - b_{2i}^{\dagger}b_{2i} ) \Bigr],
\label{eq:su2_sb_representation}
\eqa
where $\lambda_{i}^{(1),(2),(3)}$ are the real Lagrangian multipliers to enforce the local single occupancy constraint in the SU(2) slave-boson representation\cite{WEN}.
It is noted that the inclusion of the charge-charge interaction $-\frac{J}{4} n_i n_j$ in the Heisenberg Hamiltonian results in the coupling between the spin (spinon) and charge (holon) degrees of freedom.
It is easy to show that under inverse SU(2) transformation of the above expression we can readily recover the original form of the U(1) slave-boson representation (see \cite{LEE} for details).

After relevant Hubbard-Stratonovich transformations\cite{LEE}, the holon action is obtained to be 
\bqa
&& S^b({\bf A},\chi, \Delta^f, \Delta^B, h) = \int_0^\beta d\tau \Big[
\sum_i h^\dagger({\bf r}_i,\tau) ( \partial_\tau - \mu ) h({\bf r}_i,\tau) \nn
&& + \frac{J}{2} \sum_{<i,j>} |\Delta^f_{ij}|^2 \Big( \sum_{\alpha,\beta} |\Delta^b_{ij;\alpha \beta}|^2 + x^2 \Big) \nn
&& -\frac{t}{2} \sum_{<i,j>} \Big( e^{i A_{ij}} h^\dagger({\bf r}_i, \tau) U^b_{ij} h({\bf r}_j, \tau) + c.c. \Big) \nn
&& -  \frac{J}{2} \sum_{<i,j>} \Big( |\Delta^f_{ij}|^2 h^\dagger({\bf r}_i, \tau) \Delta^B_{ij} (h^\dagger({\bf r}_j, \tau))^T + c.c. \Big)
\Big],
\label{eq:su2_holon_action}
\eqa
and the spinon action,
\bqa
&& S^f(\chi, \Delta^f, \psi) = \int_0^\beta d\tau \Big[
\sum_i \psi^\dagger({\bf r}_i, \tau)  \partial_\tau  \psi({\bf r}_i, \tau) \nn
&&  + \frac{J(1- x^2)}{2} \sum_{<i,j>} \Big( |\Delta^f_{ij}|^2 + \frac{1}{2}|\chi_{ij}|^2 \Big) \nn
&& -\frac{J}{4}(1- x )^2 \sum_{<i,j>} \Big( \psi^\dagger({\bf r}_i, \tau) U^f_{ij} \psi({\bf r}_j, \tau) + c.c. \Big)
\Big].
\label{eq:su2_spinon_action}
\eqa
Here various symbol definitions are as follows.
$h({\bf r}_i,\tau) = \left( \begin{array}{c} b_{1}({\bf r}_i,\tau) \\ b_{2}({\bf r}_i,\tau) \end{array} \right)$ is the SU(2) doublet of holon field, and $\psi({\bf r}_i, \tau) = \left( \begin{array}{c} f_{1}({\bf r}_i,\tau) \\ f_{2}^\dagger({\bf r}_i,\tau) \end{array} \right)$, the SU(2) doublet of spinon field.
${\bf A}$ is the electromagnetic(EM) field.
$ U^b_{i,j}  =  \left( \begin{array}{cc} \chi^*_{ij} & - \Delta^f_{ij} \\
                           - \Delta^{f*}_{ij} & -\chi_{ij}
                         \end{array} \right)$ and
$ U^f_{i,j}  =  \left( \begin{array}{cc} \chi^*_{ij} & - 2\Delta^f_{ij} \\
                           - 2\Delta^{f*}_{ij} & -\chi_{ij}
                         \end{array} \right)$ 
are the order parameter matrices involved with hopping($\chi_{ij}$) and spinon pairing($\Delta^f_{ij}$). 
Here $\chi_{i,i+l} = \eta e^{i \alpha_{l}({\bf r}_i) } \cos (\theta_{l}^0 + \theta_{l}({\bf r}_i)) $ and $\Delta^f_{i,i+l} = \eta e^{i \beta_l({\bf r}_i) } \sin (\theta_{l}^0 + \theta_{l}({\bf r}_i))$, with  $\eta = \sqrt{ | \chi |^2 + | \Delta^f |^2 }$ and $\theta^0_l = \pm \tan^{-1} \frac{\Delta_f}{\chi}$(the sign $+(-)$ is for the ${\bf ij}$ link parallel to $\hat x$ ($\hat y$)).
$\alpha_l({\bf r}_i)$ represents the phase fluctuations of the hopping order parameter, $\beta_l({\bf r}_i)$, the phase fluctuations of the spinon pairing order parameter, and $\theta_l({\bf r}_i)$, the relative phase fluctuations between the amplitudes of hopping and spinon pairing order parameters.
$ \Delta^B_{ij}  =  \left( \begin{array}{cc} \Delta^b_{ij;11} &  \Delta^b_{ij;12} \\
                            \Delta^{b}_{ij;21} & \Delta^b_{ij;22}
                         \end{array} \right)$
is the matrix of the holon pairing order parameter.

After integration over the holon and spinon fields, we obtain the total free energy,
\bqa
F({\bf A}) = -\frac{1}{\beta} \ln \int D\chi D\Delta^f D\Delta^B \nn
e^{-\beta (F^b({\bf A}, \chi, \Delta^f, \Delta^B) + F^f(\chi, \Delta^f)) }, 
\label{eq:free_energy_su2_2}
\eqa
where $F^b({\bf A}, \chi, \Delta^f, \Delta^B ) = -\frac{1}{\beta}\ln \int Dh e^{-S^b({\bf A},\chi, \Delta^f, \Delta^B, h)}$ is the holon free energy and $F^f(\chi, \Delta^f) = -\frac{1}{\beta}\ln \int D\psi e^{-S^f(\chi, \Delta^f, \psi)}$, the spinon free energy.

The saddle point equations are solved for the spinon pairing order parameter $\Delta_f$ and holon pairing order parameter $\Delta_b$ including the hopping order parameter as a function of doping and temperature.
In Fig. 1 we display the temperature dependence of the order parameters $\Delta_f$ and $\Delta_b$ at $x = 0.1$ for $J/t = 0.2$, $0.3
$ and $0.4$.
The spin gap temperature $T^*$ and the superconducting temperature $T_c$ are identified as the temperatures below which the spinon pairing and holon pairing gaps respectively begins to open up.
Both the spinon pairing and the holon pairing order parameters are shown to change continuously at the critical temperatures, $T^*$ and $T_c$.
In Fig. 2 we show the doping dependence of $T_c$ and $T^*$ for $J/t = 0.2$, $0.3$ and $0.4$.
The spin gap temperature $T^*$ is predicted to increase with increasing antiferromagnetic coupling $J$ as expected.
It is noted that $T_c$ also increases with increasing $J$.
This is because the Cooper pairs as composites of the holon pair and the spinon pair are bose condensed as a result of coupling of the holon pair to the spinon pair whose order parameter strength increases with $J$, as can be understood from the last term in Eq. (\ref{eq:su2_holon_action}).
This is in sharp contrast to the single holon slave-boson theories\cite{KOTLIAR,FUKUYAMA,UBBENS,WEN} where $T_c$ scales only with $t$, but not with $J$.
In Fig. 3 the doping dependence of the spin gap 
\bq
\Delta_{0}  =  J (1-x)^2 \Delta_f (\cos k_x - \cos k_y)
\label{eq:spinon_energy_gap}
\eq
 at ${\bf k} = (\pi,0)$ and $T=0$ is shown for the choice of $J/t = 0.2$, $0.3$ and $0.4$.
The spin gap monotonically decreases with increasing hole concentration.
This is because the spin (spinon pair) singlet bonds are more readily broken by mobile holes with increasing vacant sites.

\section{Doping and temperature dependence of superfluid weight} 

To compute the EM current response function, we first obtain the total free energy by integrating out the three phase fields $\alpha$, $\beta$ and $\theta$ above.
We obtain a formula for the EM current response function, up to the second order, 
\begin{widetext}
\bqa
\Pi_{lm}(\omega, {\bf q}) & = & \Pi_{lm}^{b(A, A)}(\omega, {\bf q}) \nn
& - & \sum_{a^1,a^2=\theta, \alpha, \beta} \sum_{l^{'},m^{'}=\hat x,\hat y} \Pi_{ll^{'}}^{b(A,a^1)}(\omega, {\bf q}) \left[ \Pi^{b}(\omega, {\bf q}) + \Pi^{f}(\omega, {\bf q}) \right]^{-1}_{a^1,l^{'};a^2,m^{'}}  \Pi_{m^{'}m}^{b(a^2,A)}(\omega, {\bf q}),
\label{eq:ioffe_larkin_su2}
\eqa
\end{widetext}
where $\left[ \Pi^{b}(\omega, {\bf q}) + \Pi^{f}(\omega, {\bf q}) \right]^{-1}_{a^1,l^{'};a^2,m^{'}}$ is the inverse matrix element of $\Big[ \Pi^{b(a^1,a^2)}_{l^{'}m^{'}}(\omega, {\bf q})$ $+$ $ \Pi^{f(a^1,a^2)}_{l^{'}m^{'}}(\omega, {\bf q}) \Big]$.
Here $\Pi_{lm}^{b(a^1,a^2)}$ and $\Pi_{lm}^{f(a^1,a^2)}$ represent the isospin current response functions of holons and spinons respectively to gauge fields $a^1, a^2 = \theta, \alpha,$ or $\beta$.
Similarly, $\Pi_{lm}^{b(A, A)}(\omega, {\bf q})$ is the EM current response function of holon to the EM field.
$\Pi_{lm}^{b(a^1,A)}$ is the isospin current response function of holon to both gauge fields $a$ and $A$.
It is noted that the SU(2) isospin current of holon(spinon) is defined as $j_{\alpha,l}^{b(f)} = -\beta \frac{\delta F^{b(f)}}{\delta a^\alpha_{i,i+l}}$,  and the `EM' current, $j_l = -\beta \frac{\delta F}{\delta A_l}$.
In the SU(2) slave-boson theory, the response function $\Pi_{ll^{'}}^{b(A,a^1)}$ of the holon isospin current to the EM field vanishes owing to the contribution of the $b_2$-boson in the static and long-wavelength limit\cite{WEN}.
Therefore the superfluid weight of the total system is given only by the holon current response function,
\bqa
\frac{n_s}{m^*} = - \frac{1}{e^2} \lim_{ {\bf q} \rightarrow 0} \Pi_{ll}^{b(A, A)}(\omega=0, {\bf q} ).
\label{eq:su2_superfluie_weight}
\eqa
Here $\Pi_{lm}^{b(A, A)}(\omega, {\bf q})$ is computed from the use of the usual linear response theory for the holon action Eq.(\ref{eq:su2_holon_action})\cite{SCHRIEFFER}.

In Fig. 4(a), with the use of our slave-boson theory\cite{LEE} we show the temperature dependence of superfluid weight for a wide range of doping concentrations covering both underdoped and overdoped regions.
The slope of each curve represents the variation of superfluid weight with temperature which is only in qualitative (but not quantitative) agreement with observations\cite{SHENG,PANA0,PANA} in that the superfluid weight $\frac{n_s}{m^*}(x,0)$ at $T = 0 K$ is overestimated and further $\frac{n_s}{m^*} (x,T)$ reaches zero rather abruptly compared to the observation which showed a slower drop\cite{PANA0}.
However a close look at Fig. 4 for comparison with experiment\cite{PANA0} reveals a linear slope behavior only in the lower temperature region and a quick non-linear drop of superfluid weight at the higher temperature region.
Such a slope variation is in good agreement with our results although our theory predicted a faster drop in $\frac{n_s}{m^*}(T)$.
The continuous drop in the predicted superfluid weight with temperature indicates a second order phase transition.
We believe that improvement can be achieved if we can introduce the $t^{'}$ term in our t-J Hamiltonian which is known to cause a good agreement with the observed dispersion energy by showing realistic cold and hot spots.
The contribution of the hot spot in the Brillouin zone becomes increasingly important with increasing temperature for classical phase fluctuations to occur at $T_c$.

As temperature $T$ increases, the predicted superfluid weight shows a tendency of linear decrease in $T$ with nearly identical slopes particularly at low temperatures in the underdoped region.
To rigorously check such a propensity of doping independence in the slope of $\frac{n_s}{m^*}$ vs. $T$, we computed differences in the superfluid weight at $T=0$ and $T \neq 0$, that is, $\frac{n_s}{m^*}(x,T) - \frac{n_s}{m^*}(x,0)$ at each $x$.
The results are shown in Fig. 4 (b).
A decrease of the superfluid weight with temperature at all hole concentrations is predicted although the temperature dependence of the superfluid weight does not show a clear linearity at all temperatures as shown in Fig. 4 (b).
This figure displays that in agreement with observation there is a tendency of linear decrease with a nearly identical slope particularly in the underdoped region only at low temperatures.
Although the temperature dependence of $\frac{n_s}{m^*}(x,T)$ does not show a global linearity by yielding a good fit to the empirical relation of $\frac{n_s}{m^*}(x,T) = \frac{n_s}{m^*}(x,0) - \alpha(x) T - \alpha^{'}(x) T^2 - O(T^3)$, the linear slope $\alpha$ is predicted to be doping independent only in the underdoped region and at low temperatures as is clearly manifested in Fig. 4 (b).
As shown in the figure it is found that the variation of the slope, that is, the superfluid weight $n_s/m^*$ vs. $T/t$ at low temperatures does not appreciably change in the underdoped region ($x=0.04$, $0.07$ and $0.1$).
On the other hand, in the overdoped region ($x=0.16$, $0.18$ and $0.2$) the slope of the superfluid weight with temperature is no longer doping-independent quite unlike the underdoped case, again, in agreement with observation.
The predicted optimal doping is $x_o = 0.13$.
We find that there exists crossings of superfluid weights between the underdoped and overdoped cases in the plane of $n_s/m^*$ vs. $T/t$ as shown in Fig. 4(a).
This prediction is also in agreement with measurements\cite{SHENG,PANA0,PANA}.

In Fig. 4(c) we display the the normalized superfluid weight ($\frac{n_s/m^*(x,T)}{n_s/m^*(x,0)}$) as a function of the normalized temperature ($T/T_c$) for various hole concentrations.
As shown in the figure, the normalized superfluid weight begins to decrease linearly at low temperature and drops more rapidly at higher temperature, by showing a convex shape in agreement with experiments\cite{SHENG,PANA0,PANA}.
The predicted decrease of superfluid weight with temperature occurs as a consequence of diminishing spin singlet pairing order with increasing temperature.
This is readily understood from the holon pairing term (the last term in Eq. (2)) which reveals the coupling of the spin (spinon) singlet pairing order with the holon pairing.
Consequently the decrease of the superfluid weight with increasing temperature is caused by breaking the bond of spinon singlet pairs coupled to the holon pairs, whose spin bond strength decreases as temperature increases.
Although the superfluid weight shows a linear decrease in agreement with observation, the coefficient $\beta$ in the relation of $\frac{n_s/m^*(x,T)}{n_s/m^*(x,0)} = 1 - \beta (\frac{T}{T_c})$ is scattered around $0.055$ and thus fails to quantitatively agree with the empirical value of $\beta \approx 0.5$.
However the important finding here is the $\beta$ is independent of doping in the underdoped region\cite{PANA}.
The quantitative discrepancy indicates that the inclusion of the next nearest hopping ($t^{'}$) term is necessary.
This is because the Fermi surface segment grows from the nodal point initially at half-filling and closes near $(\pi, 0)$ as hole doping reaches optimal doping and beyond as is well-known from the angle resolved photoemission spectroscopy measurements\cite{MARSHALL}.
In the future the inclusion of the $t^{'}$ term will have to be made to take a good account of classical phase fluctuations in the region of the hot spot at high temperatures at and near $T_c$

According to the present theory the superconductivity arises by the condensation of the Cooper pairs of d-wave symmetry as composites of the holon pairs of s-wave symmetry and the spinon pairs of d-wave symmetry, but not by the single-holon bose condensation unlike other theories.
To put it otherwise the Cooper pairs can be regarded as composite objects resulting from such coupling of the spin (spinon) and charge (holon) degrees of freedom.

To clearly show the effect of coupling between the spin(spinon) and charge(holon) degrees of freedom, we calculated the superfluid weight by varying the antiferromagnetic spin-spin coupling, that is, the Heisenberg coupling $J$.
In Fig.5, we show the $J$ dependence of superfluid weight at $x=0.1$ for three different choices of $J/t=0.2, 0.3$ and $0.4$. 
As $J$ increases, a decreasing tendency in the slope of the superfluid weight vs. temperature is predicted.
For large antiferromagnetic coupling $J$ and thus the strong spinon pairing bond (strength), it is more difficult to thermally break the holon-pairs because of the increased holon-spinon coupling.
It is again reminded that the Cooper pair is obviously a composite of the spinon pair and the holon pairs which are coupled to each other as is shown in the last term of Eq.(\ref{eq:su2_holon_action}).
This results in the more slowly decreasing slope of the superfluid weight with increasing temperature for large $J$, as is shown in the figure.

In Fig.6, we show the predicted boomerang behavior in the plane of $T_c$ vs. $\frac{n_s}{m^*}(x,T \rightarrow 0)$, by showing a linear relationship between the two at low hole concentrations in the underdoped region and the `reflex' behavior in the overdoped region based on the U(1) theory.
By the reflex behavior we mean the decrease of both $T_c$ and $\frac{n_s}{m^*}$ as the hole doping concentration increases in the overdoped region.
This predicted boomerang behavior is consistent with the measurements\cite{UEMURA,UEMURA_NATURE,BERNHARD} of the muon-spin-relaxation rates $\sigma$ ($\sigma \propto n_s/m^*$).
Here we focus our attention to its cause.
The pseudogap (spin gap) temperature decreases in the overdoped region in agreement with observation.
The predicted pseudogap (spin gap) caused by the appearance of the spinon pairing order $\Delta^f$ decreases in the overdoped region and this, in turn, causes a decrease in holon pairing $\Delta^b$ owing to its coupling to the spinon pairing order whose strength decreases with increasing hole concentration. 
As a consequence, $\frac{n_s}{m^*}$ has to show the boomerang behavior in the plane of $\frac{n_s}{m^*}$ vs. $T_c$.
The boomerang behavior of the SU(2) theory is found to be qualitatively the same as that of the U(1) theory.

\section{Universal scaling behavior of pseudogap based on the holon-pair bose condensation theory} 

The single particle(electron) propagator of interest is given by a convolution integral of spinon and holon propagators in the momentum space\cite{WEN},
\bqa
G_{ \alpha \beta}({\bf k},\omega) & = & \frac{i}{2} \int \frac{d{\bf k}^{'} d\omega^{'}}{(2\pi)^3} \Bigl[ \sum_{l,m}G^f_{ \alpha \beta l m }({\bf k}+{\bf k}^{'},\omega+\omega^{'}) \times \nn
&& G^b_{ml}({\bf k}^{'},\omega^{'}) \Bigr].
\label{eq:convolution}
\eqa
The mean field Green's functions are $G^f_{\alpha\beta l m}({\bf k},\omega) = -i\int dt \sum_x e^{i\omega t-i {\bf k} \cdot {\bf x}} < T [ \psi_{\alpha l}({\bf x},t) \psi_{\beta m}^\dagger(0,0) ] >$
and $G^b_{lm}({\bf k},\omega) = -i\int dt \sum_x e^{i\omega t-i {\bf k} \cdot {\bf x}} < T [ b_{l}({\bf x},t) b_{m}^\dagger(0,0) ] >$ respectively
(The symbol $<$ $>$ here refers to the finite temperature  ensemble average of an observable quantity $O$, $<O> \equiv \frac{1}{Z}{\rm tr}(e^{-\beta H}O)$).

The one electron removal spectral function, $I({\bf k}, \omega )$ is obtained from \cite{RANDERIA},
\bqa
I({\bf k},\omega) = -\frac{1}{\pi}{\rm Im G}({\bf k},\omega + i 0^+) f( \omega ), 
\eqa
where $f(\omega)$ is the Fermi distribution function.
The Heisenberg coupling constant of $J=0.2$ $t$ and the hopping strength of $t=0.44$ $eV$\cite{HYBERTSEN} are chosen in the present calculations.
Using the SU(2) theory, the predicted values of optimal hole doping $x_o$, pseudogap temperature $T^*$ and bose condensation temperature $T_c$ are $x_o = 0.13$, $T^* = 0.029 t$($148 K$) and $T_c = 0.021 t$ ($107.2 K$) respectively.
For the evaluation of the spectral functions $I({\bf k}, \omega)$, the line width of the Lorentzian function is set to be $\epsilon = 0.01 t (4.4 meV)$ with the choice of $t=0.44 eV$.

It is known from the ARPES measurements of LSCO\cite{INO} and BSCCO\cite{DING97} that a good Fermi nesting appears in a region near the $M$ point (${\bf k}=(\pi,0)$) at which maximal antiferromagnetic correlations between electrons are realized.
The predicted binding energy monotonically decreases with the increase of momentum ${\bf k}$ along $\Gamma - M$, in agreement with the ARPES\cite{CAMPUZANO99}.

In Fig.7 we display the computed spectral functions $I({\bf k},\omega)$ at ${\bf k}=(\pi,0)$ both below and above $T_c$.
The computed gap size at ${\bf k}=(\pi,0)$ is predicted to coincide with the spin gap size $\Delta_0$ in Eq.(\ref{eq:spinon_energy_gap}).
The peak position occurs at smaller binding energies at temperatures ($0.022t$($112.3 K$)) above $T_c$. 
As is shown in Fig. 7(a), a spin gap begins to open at a higher critical temperature $T^* = 0.029t$($147.9 K$) above $T_c$.
The predicted peak position shift, that is, the shift of leading edge gap is much slower below $T_c$.
This trend is in complete agreement with observations\cite{CAMPUZANO99,DING}.
Recently ARPES measurements\cite{FEDOROV} for $Bi_2 Sr_2 Ca Cu_2 O_8$ showed that there exists a quasiparticle peak above $T_c$ and the peak intensity decreases with increasing temperature, while the enhancement of peak intensity occurs markedly at temperatures below $T_c$.
Our computed results revealed the presence of the quasiparticle peaks in both below and above $T_c$ as shown in Fig. 7 (a).
A recent ARPES study of Feng et al.\cite{FENG} claimed that the enhancement of the quasiparticle peak below $T_c$ is caused by bose condensation.
Our theory proves that such enhancement of the quasiparticle peak below $T_c$ is caused by the bose condensation.
For this verification, in our calculation we first removed completely the contribution of the holon-pair channels (which comes from momenta ${\bf k}^{'}=(0,0)$ and ${\bf k}^{'}=( \pi,\pi)$). 
The sharp peak is now seen to disappear and only the hump structure is predicted at all temperatures at and below $T_c$ while the pseudogap size remains unchanged.
This clearly indicates that the pseudogap is caused by the formation of spin singlet pairs and that the presence of holon-pair bosons is essential for yielding the observed enhancement of quasiparticle peaks at the $M$ point below $T_c$. 
In short, the quasiparticle peak is caused by bose condensation below $T_c$. 
We may now argue that the appearance of the hump in the absence of bose condensation is seen to be caused by the antiferromagnetic spin fluctuations of containing the shortest possible correlation length, that is, the spin singlet pair excitations.
Both the SU(2) and U(1) slave-boson theories were found to predict the enhancement of quasiparticle peaks at ${\bf k} = (\pi, 0)$ and ${\bf k} = (\pi/2, \pi/2)$ below $T_c$.
As is shown in Fig. 7, the predicted leading edge gap at ${\bf k}=(\pi,0)$ showed a continuous increase as temperature decreases from a pseudogap temperature $T^*$ to temperatures below the superconducting temperature $T_c$. 
This indicates that the observed leading edge gap with the appearance of sharp quasiparticle peak below the superconducting transition temperature is originated from the persistence of the spin gap with the emergence of bose condensation (holon pair bose condensation).
This implies that the observed quasiparticle peak in the superconducting state is attributed to the presence of the coupling of the spin (spinon pairing order $\Delta^f$) to the charge (holon pairing order $\Delta^b$) degrees of freedom as shown in the last term of Eq.(\ref{eq:su2_holon_action}).

In Fig. 8 the doping dependence of spectral functions is displayed for underdoped, optimally doped and overdoped cases at a low temperature, $T = 0.004 t$ ($T = 20.4 K$) below $T_c$, near the $M$ point (${\bf k} = (0. 8 \pi, 0)$) for comparison with observation.  
In agreement with the ARPES\cite{SHEN}\cite{HARRIS} the predicted spectral weight of the sharp quasiparticle peaks is seen to increase as the hole concentration increases upto a tested value in the overdoped region. 
The leading edge gap is shown to decrease with increasing hole concentration in agreement with observations\cite{INO}.
In our calculations the dip in the peak-dip-hump structure is not predicted.
It is claimed that the dip is caused by electron-phonon coupling\cite{LANZARA}, which is not considered in our present treatment of the t-J Hamiltonian.  

Figs. 9.(a) displays as a function of doping rate the ratios of the pseudogap $2 \Delta_0(k=(\pi,0))$ at $0 K$ to the superconducting temperature $T_c$ and the spin gap temperature $T^*$, that is, $\frac{2\Delta_0}{k_B T_c}$ and $\frac{2\Delta_0}{k_B T^*}$ respectively.  
It is possible that each high $T_c$ cuprate(e.g., LSCO, YBCO and BSCCO) may have an effectively different $J$ value causing variation in $T_c$.
It is quite encouraging to find that in agreement with observations\cite{ODA}, the ratio $\frac{2 \Delta_0(k=(\pi,0))}{k_B T_c}$ decreases rapidly with hole concentration $x$, showing a $J$ independent and thus sample-independent universal scaling behavior with $x$.
On the other hand, $\frac{2 \Delta_0(k=(\pi,0))}{k_B T^*}$ shows a nearly doping independence of ranging its value between $4$ and $6$, as shown in Fig. 9.(b).
The doping independence of $\frac{2 \Delta_0(k=(\pi,0))}{k_B T^*}$ is due to the fact that both $\Delta_0$ and $T^*$ simultaneously decreases with $x$ in a concerted manner as is shown in Figs. 2 and 3.
That is, the ratio $\frac{2 \Delta_0(k=(\pi,0))}{k_B T^*}$ should remain unchanged by revealing that both $\Delta_0$ and $T^*$ decrease at an equal rate with increasing hole concentration $x$.
On the other hand, the non-monotonously decreasing behavior of $\frac{2 \Delta_0(k= (\pi,0))}{k_B T_c}$ is attributed to the arch shape of $T_c$ line as a function of hole concentration\cite{LEE}.
The arch shape bose condensation temperature is due to the coupling of the holon pairs and spinon pairs.
In the underdoped region, spinon pairing order $\Delta^f$ is large and $T_c$ increases with increasing $x$.
On the other hand, in the overdoped region $\Delta^f$ continuously decreases with increasing $x$.
In this region $T_c$ also decreases with increasing $x$ as a result of the diminishing nature of $\Delta^f$.
It is found that $\frac{2\Delta_0}{k_B T_c}$ and $\frac{2\Delta_0}{k_B T^*}$ do not appreciably change with the variation of Heisenberg coupling $J$.
This is due to the nature that $\Delta_0$, $T^*$ and $T_c$ are found to be proportional to $J$, keeping these ratios unchanged\cite{LEE}.
Thus we found that independent of Heisenberg exchange (antiferromagnetic) coupling there exists a universal scaling behavior of the ratios of $\Delta_0/T_c$ and $\Delta_0/T^*$ respectively.

\section{Summary}
We stress that the Cooper pair is simply a composite of spinon pair and holon pair as a consequence of the mutual coupling in the slave-boson language. 
The spinon pair (spin singlet pair) formation occurs below the spin gap (pseudogap) temperature. 
Thus it is needless to say that bose condensation or superfluidity should occurs below the spin gap temperature.
From the present study we found the correct doping and temperature dependences of both the superfluid weight and spectral function and the universal scaling behavior of pseudogap in agreements with observations.
Regarding the doping and temperature dependences of the superfluid weight, we observed the nearly doping independence in the negative slope of superfluid weight vs. temperature in the underdoped region, thus correctly reproducing the experimentally observed relation, $\frac{n_s}{m^*}(x,T) = \frac{n_s}{m^*}(x,0) - \alpha T$ with $\alpha$, independent of hole concentration.
Our results which showed a linear decrease of the superfluid weight are qualitative, but not numerically accurate as pointed out earlier.
On the other hand, in the overdoped region the predicted slope of the superfluid weight varies with both hole concentration and temperature in agreement with observation.
The boomerang behavior in the locus of both $\frac{n_s}{m^*}$ and $T_c$ with the variation of hole concentration in the plane of $\frac{n_s}{m^*}(x,T \rightarrow 0)$ vs. $T_c$  is in agreement with the $\mu$-SR experiments.
The decreasing (boomerang) behavior of $\frac{n_s}{m^*}$ and $T_c$ in the overdoped region is attributed to the weakened spinon pairing order coupled to the holon pairing order.
From the present study we find additional salient features in agreements with observations.
They are
1. the sharp quasiparticle peak observed below $T_c$ is attributed to the bose condensation, and the hump is caused by the antiferromagnetic spin fluctuations of containing the shortest possible correlation length namely the spinon (spin) singlet pair excitations, 
2. the spectral intensity of quasiparticle peak shows an increasing trend with hole concentrations in the underdoped region and the pseudogap (spin gap) size at ${\bf k}=(\pi,0)$ continuously increases as temperature decreases from the pseudogap temperature $T^*$ to temperatures at and below the superconducting temperature $T_c$,
3. a decreasing trend of spectral peak intensity is predicted as temperature increases,
and,
4. there exists a universal (i.e., sample independent) scaling behavior of both $\frac{2 \Delta_0}{k_B T_c}$ and $\frac{2 \Delta_0}{k_B T^*}$ with $x$.
Finally coupling between the spin (spinon pairing order $\Delta^f$) and the charge (holon pairing order $\Delta^b$) is important to explain the observed phase diagram, the doping and temperature dependence of the superfluid weight, its boomerang behavior, the doping and temperature dependence of the spectral function and the universal scaling behavior of the pseudogap.
We stress again the Cooper pair is a composite of the spinon pair and the holon pair to yield an object of charge $+2e$ (or $-2e$) and spin $0$.

One(SHSS) of us acknowledges the generous supports of Korea Ministry of Education(HakJin Program) and the Institute of Basic Science Research at Pohang University of Science and Technology.
He is also grateful to Chang Ryong Kim for helpful discussions.

\newpage
\centerline{FIGURE CAPTIONS}
\begin{itemize}

\item[Fig. 1 ]
Temperature dependence of the spinon pairing order parameter ($\Delta_f$) and   the holon pairing order parameter ($\Delta_b$) at $x=0.1$ for $J/t = 0.2$, $0.3$ and $0.4$.

\item[Fig. 2 ]
Doping dependence of the superconducting temperature $T_c$ and the spin gap temperature $T^*$
for $J/t = 0.2$, $0.3$ and $0.4$.

\item[Fig. 3 ]
Doping dependence of the spin gap $\Delta_0(k=(\pi,0))$ at $T=0$ for $J/t = 0.2$, $0.3$ and $0.4$.

\item[Fig. 4 ]
(a) Temperature dependence of the superfluid weight $\frac{n_s}{m^*}$ for underdoped($ x =0.04, 0.07, 0.1$), optimal doping($ x =0.13$) and overdoped($x = 0.16, 0.18, 0.2$) rates with $J/t=0.2$.  
(b) The difference of superfluid weights between $T \neq 0K$ and $T=0K$, that is, $\frac{n_s}{m^*}(T) - \frac{n_s}{m^*}(T=0)$. Each line represents a fitting to the computed superfluid weight. 
(c) The normalized superfluid weight $\frac{n_s/m^*(x,T)}{n_s/m^*(x,0)}$ as a function of scaled temperature $T/T_c$ for same doping concentrations and $J$ value as (a).

\item[Fig. 5 ]
The difference of superfluid weights between $T \neq 0K$ and $T = 0K$, $\frac{n_s}{m^*}(T) - \frac{n_s}{m^*}(T=0)$ for a underdoped hole concentration $x = 0.1$ with $J/t=0.2, 0.3$ and $0.4$.  
Each line represents a fitting to the computed superfluid weight.

\item[Fig. 6 ]
The superfluid weight $\frac{n_s}{m^*}(x,T)$ at $T=0.001t$ (equivalent to $T \sim 5K$ with the use of $t=0.44eV$\cite{HYBERTSEN}) vs. superconducting temperature $T_c$ with the choice of $J/t=0.2$ based on the U(1) theory.
The open box represents hole concentration starting from $x=0.01$ to $x=0.22$ and the arrow denotes the direction of increasing doping rate. 
The predicted optimal doping rate is $x_o = 0.07$ with the U(1) theory.
The SU(2) theory predicts qualitatively the same boomerang behavior.

\item[Fig. 7 ]
Temperature dependence of the spectral functions $I({\bf  k}, \omega)$ for ${\bf k}=(\pi,0)$ at the predicted optimal doping of $x_o = 0.13$ ($T_c=0.021t$($107.2 K$), $T^*= 0.029t$($147.9 K$)) with $J=0.2t$ for ${\bf k}=(\pi,0)$. 
(a) $I({\bf k},\omega)$ with all momenta of holon included and
(b) $I({\bf k},\omega)$ with the exclusion  of  only ${\bf k}^{'}=(0,0)$  and  $(\pi,\pi)$.

\item[Fig. 8 ]
Doping dependence of spectral functions $I({\bf k},\omega)$ near the $M$ point (${\bf k}=(0.8 \pi , 0)$) at temperature $0.004t$($20.4K$) below $T_c$ for underdoped($x=0.04$ and $x=0.085$), optimal doping($x_o =0.13$) and overdoped($x=0.2$) rates.

\item[Fig. 9 ]
The ratios of the spinon pairing gap at $T=0K$ to 
(a) the superconducting temperature, $\frac{2 \Delta_0(k=(\pi,0))}{k_B T_c}$ and
(b) the spin gap temperature, $\frac{2 \Delta_0(k=(\pi,0))}{k_B T^*}$ 
as a function of $\frac{x}{x_o}$ where $x_o = 0.13$ is the predicted optimal doping concentration for $J/t=0.2$ and $0.3$, and $x_o=0.14$ for $J/t=0.4$. 
The experimentally obtained universal ratios for $Bi_2Sr_2CaCu_2O_{8+x}$, $La_{2-x}Sr_xCuO_4$ and $Tl_2Ba_2CuO_6$ are denoted as solid circles\cite{ODA}.

\end{itemize}

\begin{widetext}

\newpage
\begin{figure}
        \includegraphics{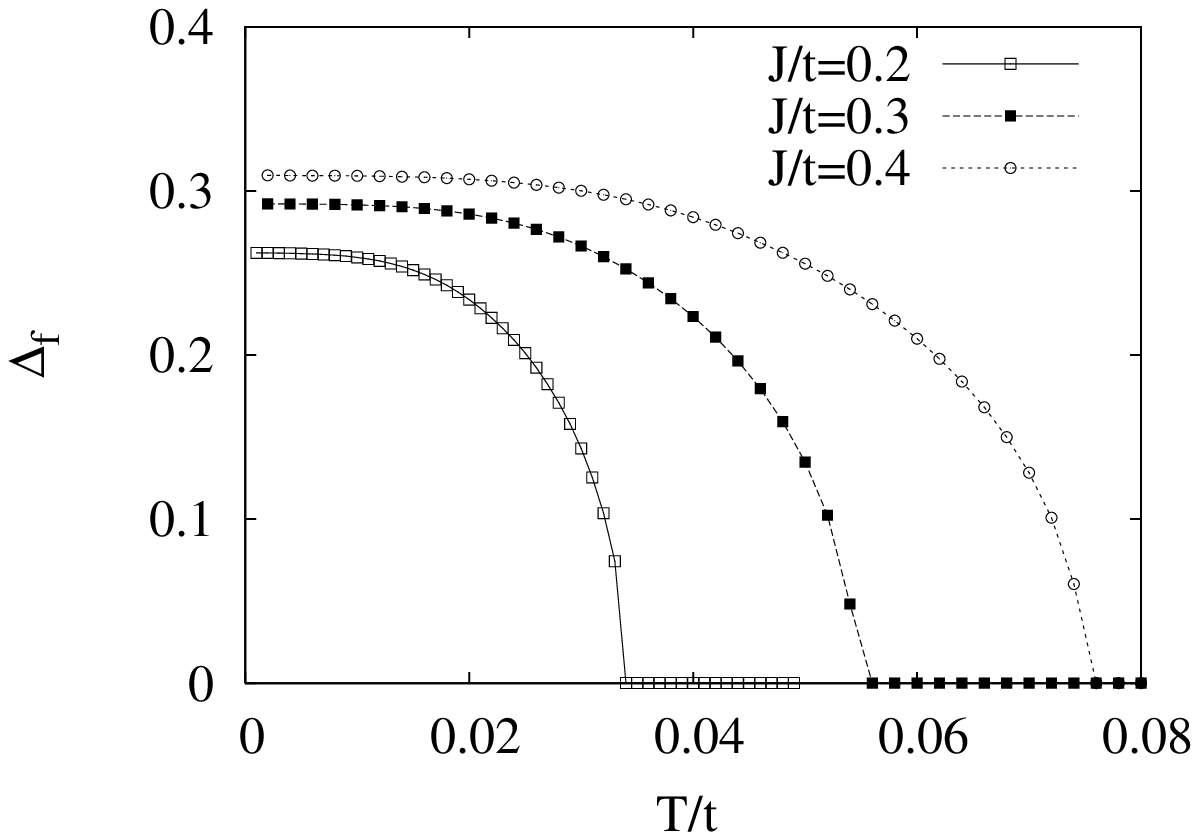}
        \includegraphics{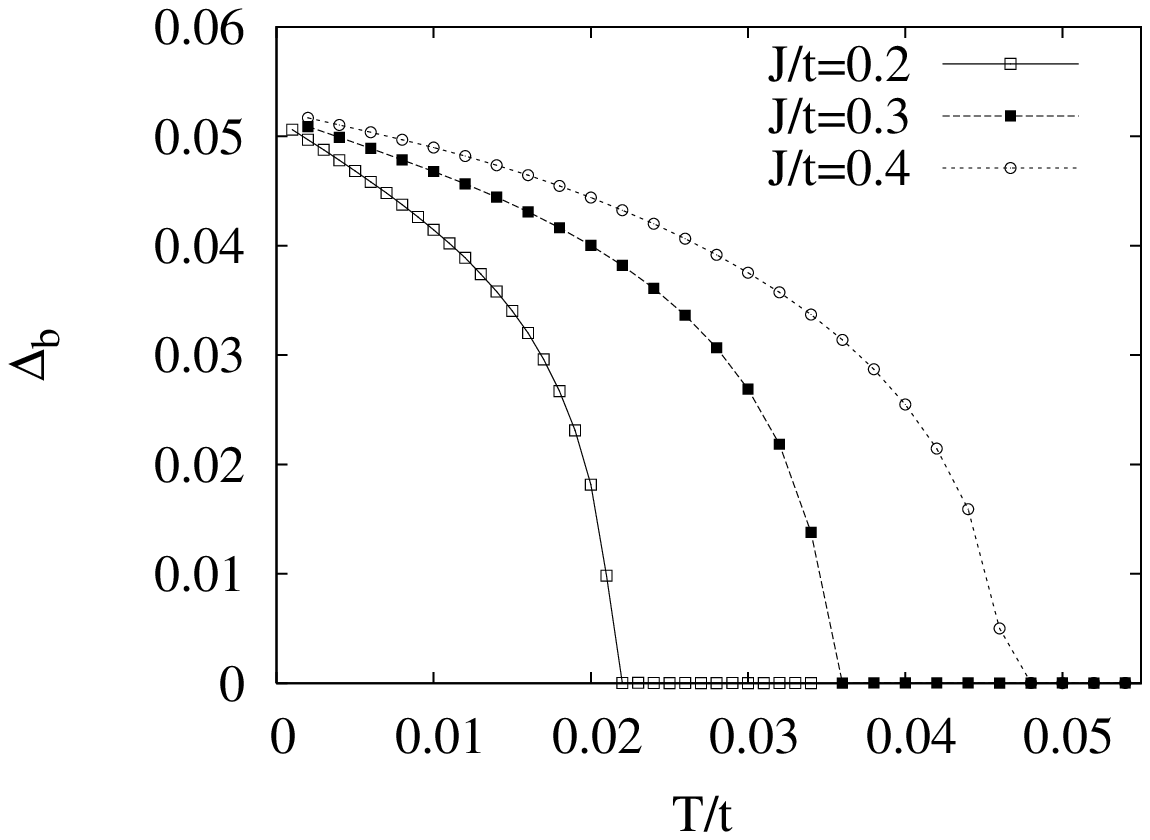}
\label{fig:1}
\caption{}
\end{figure}

\begin{figure}
        \includegraphics{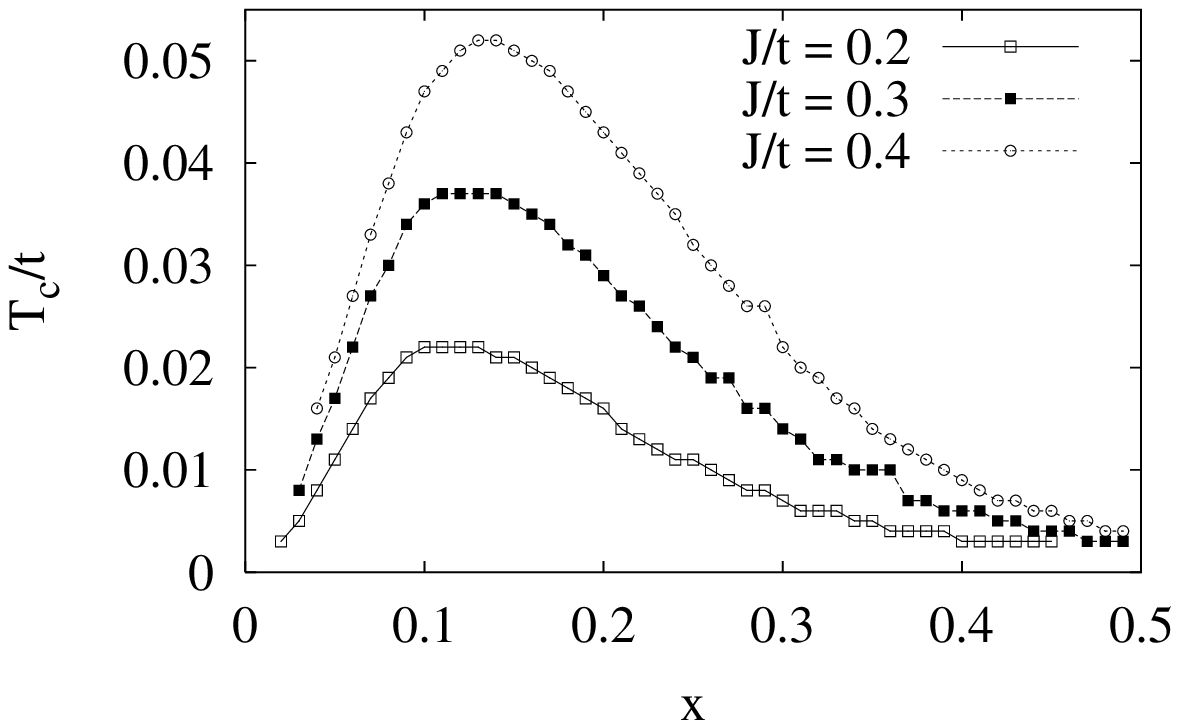}
        \includegraphics{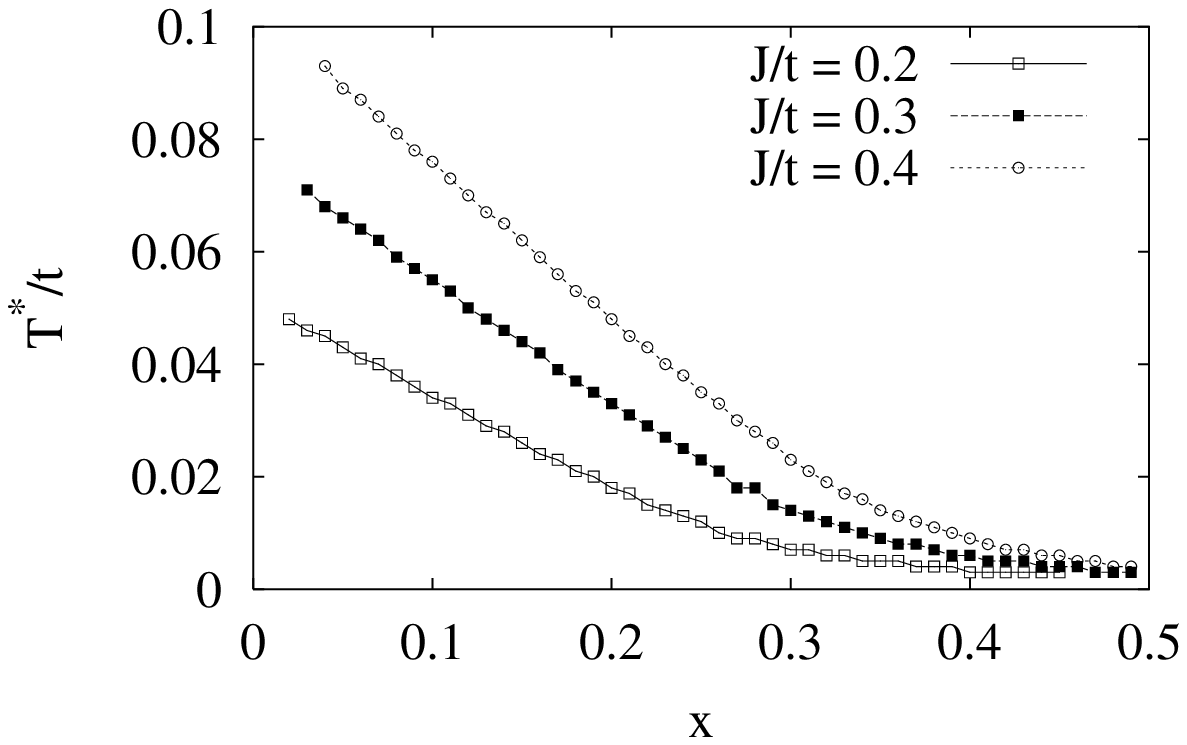}
\label{fig:2}
\caption{}
\end{figure}

\begin{figure}
        \includegraphics{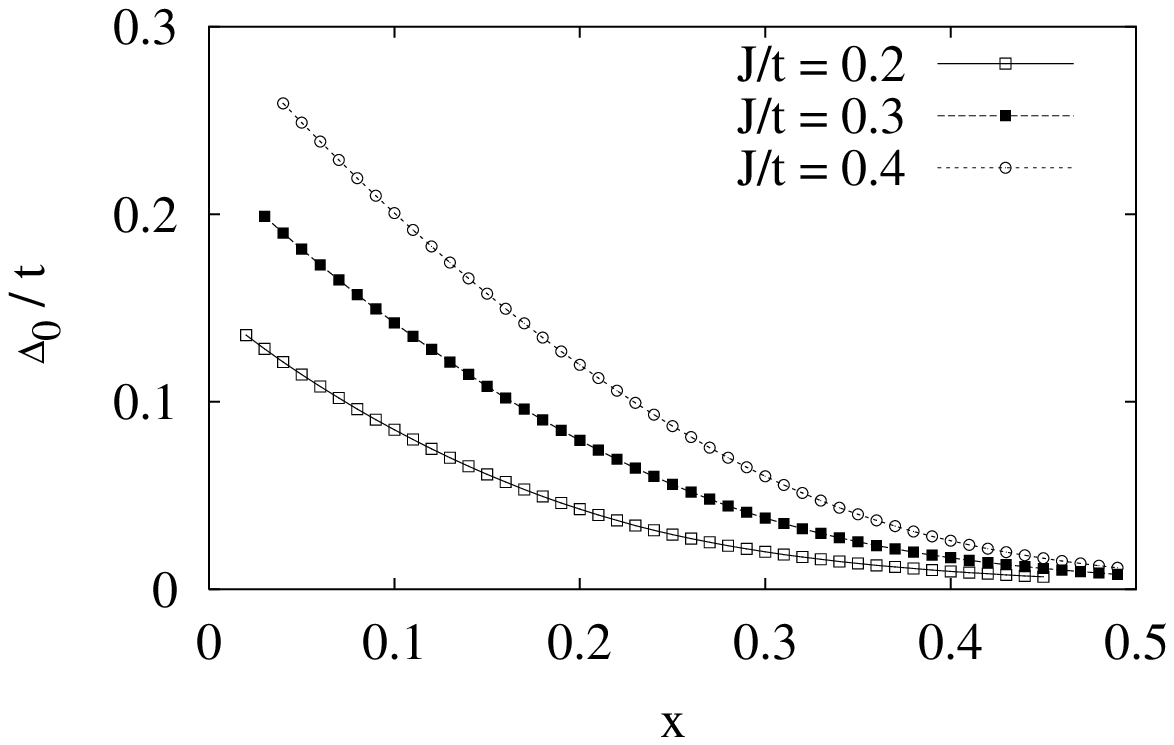}
\label{fig:3}
\caption{}
\end{figure}

\begin{figure}       
        \includegraphics[height=7cm,width=10cm]{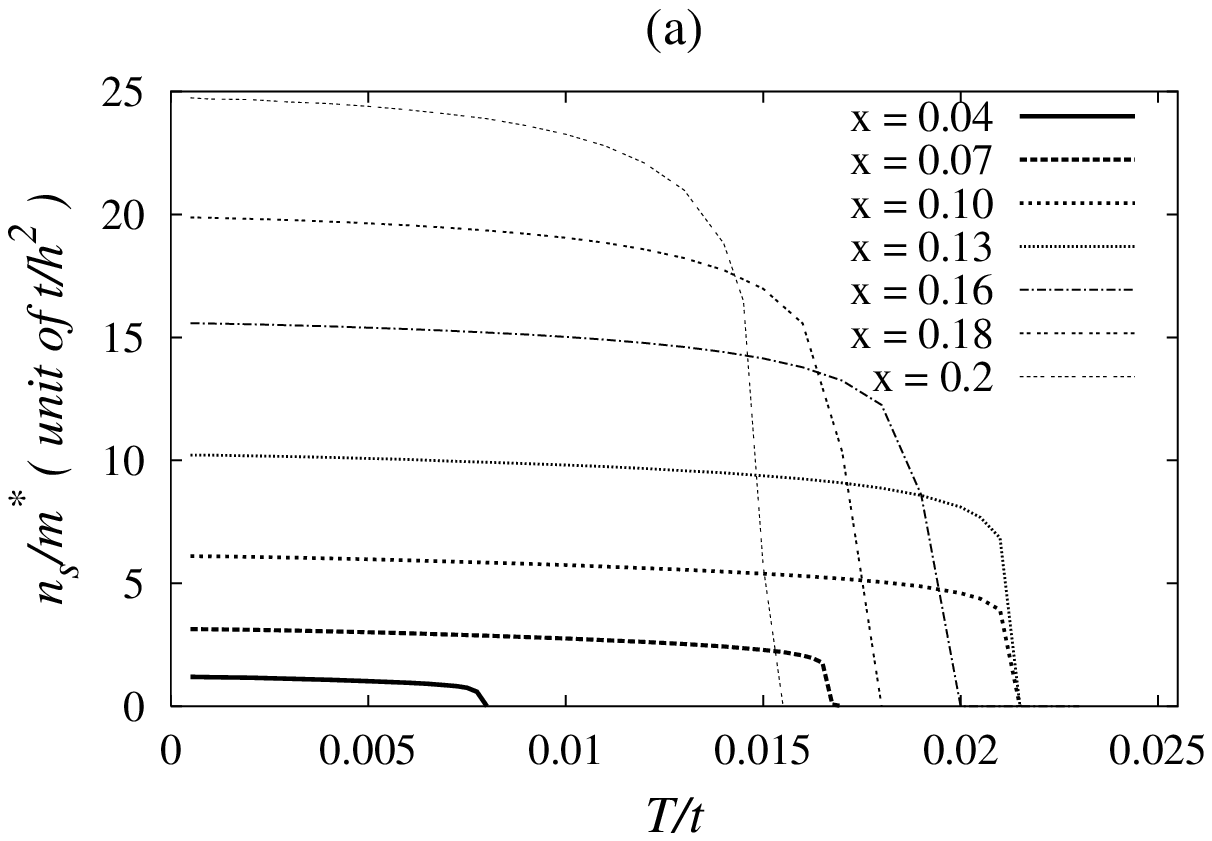}
        \includegraphics[height=7cm,width=12cm]{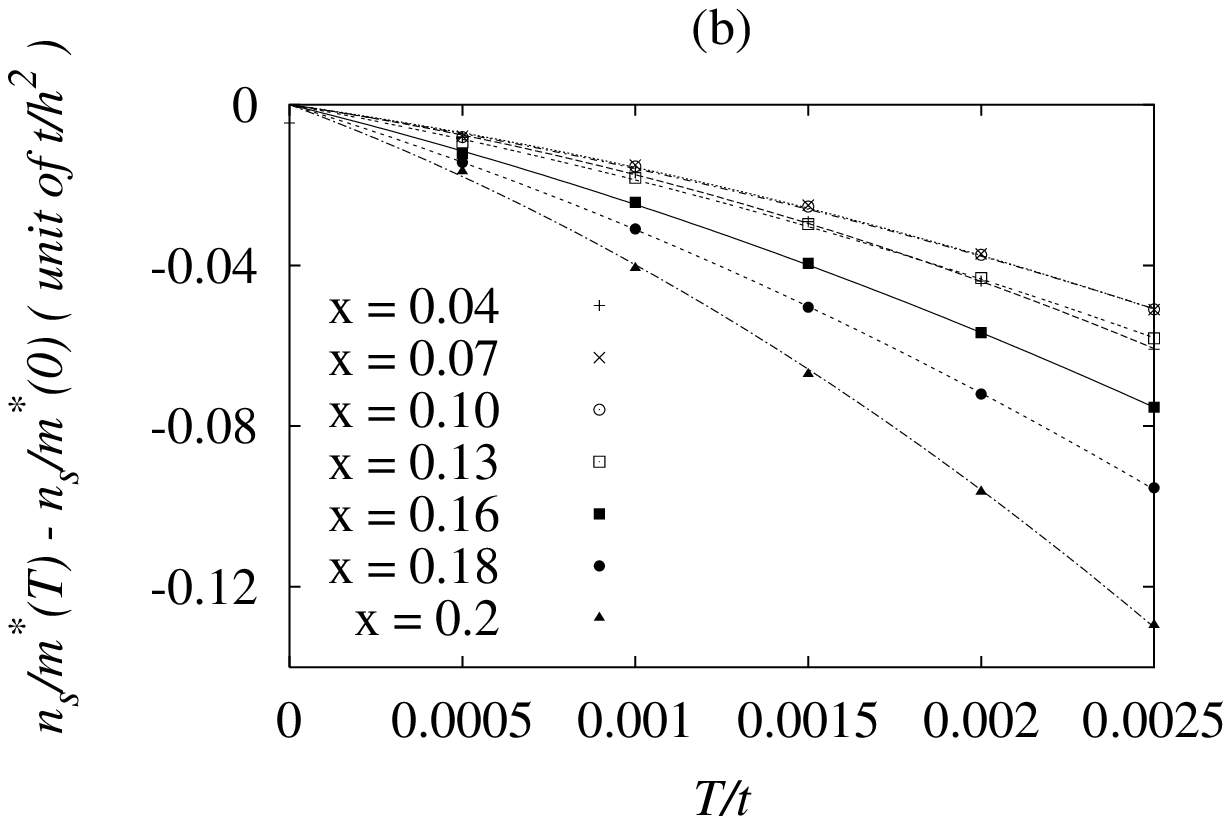}
        \includegraphics[height=7cm,width=11cm]{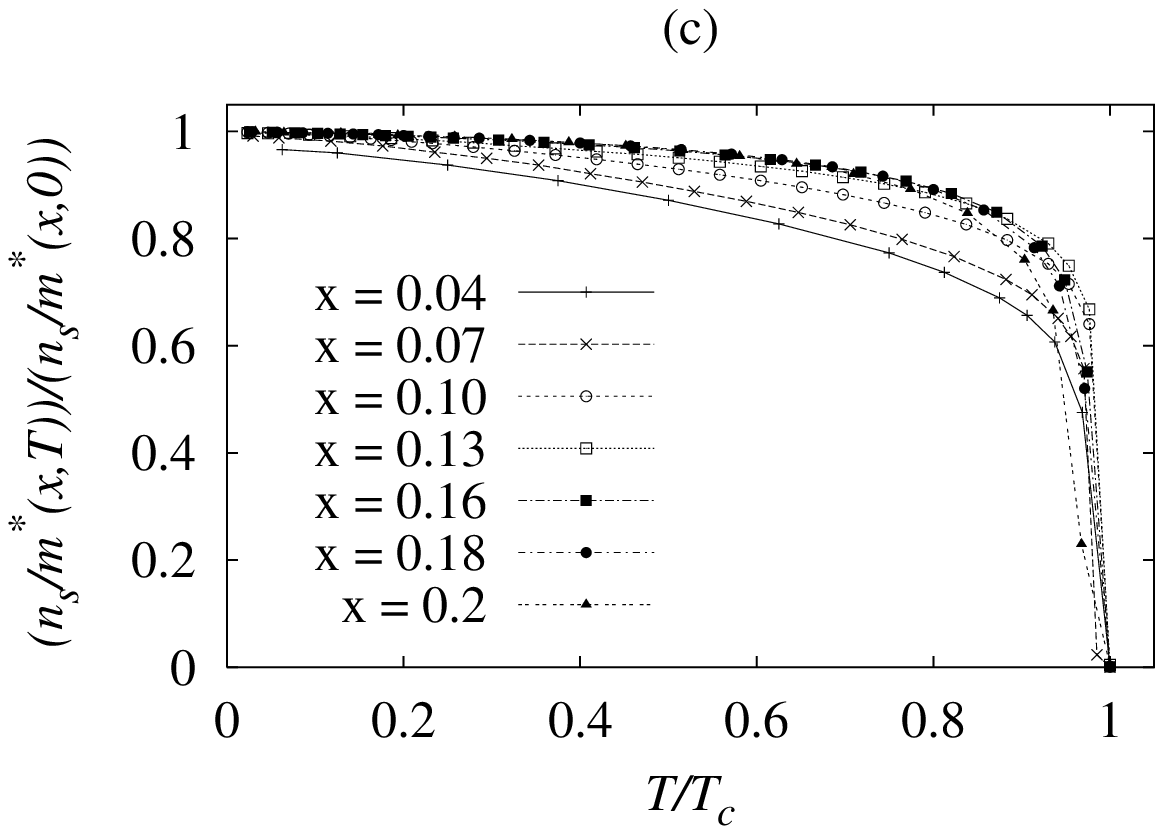}
\label{fig:4}
\caption{}
\end{figure}

\begin{figure}
        \includegraphics{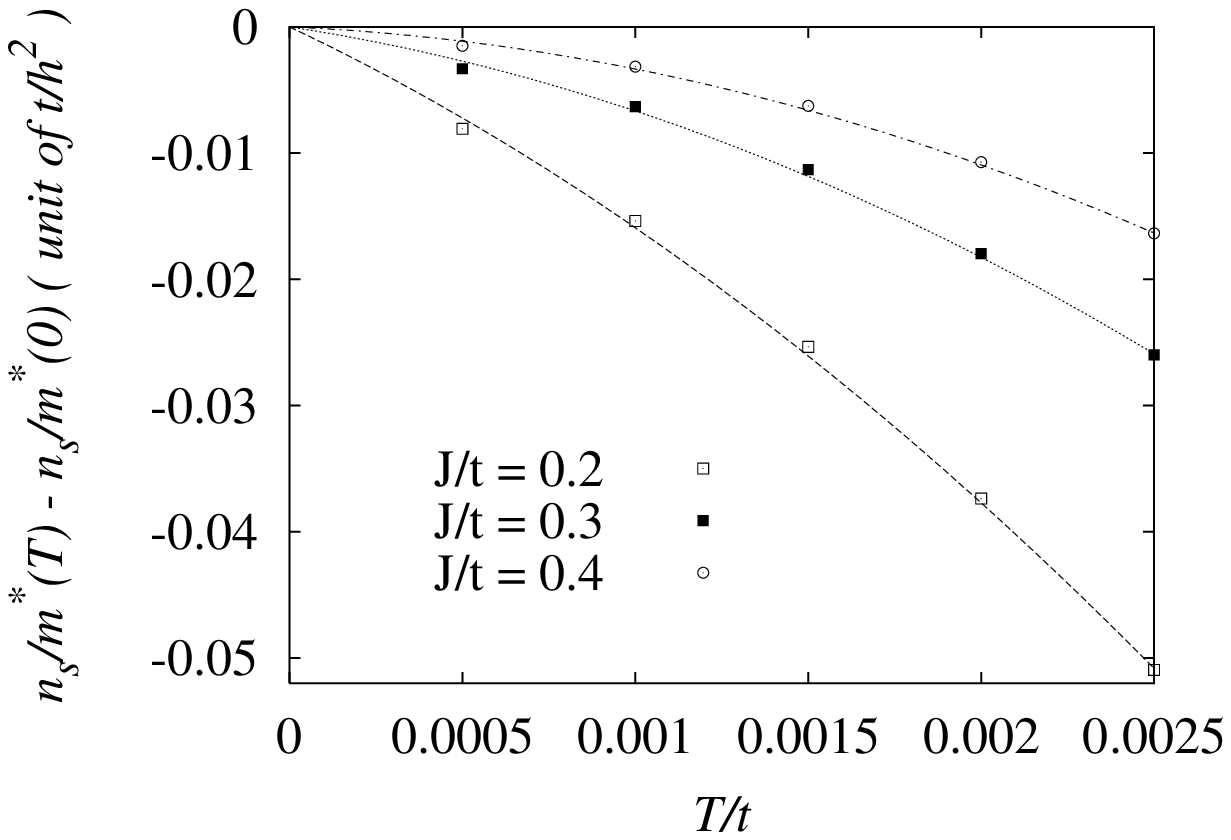}
\label{fig:5}
\caption{}
\end{figure}

\begin{figure}
        \includegraphics{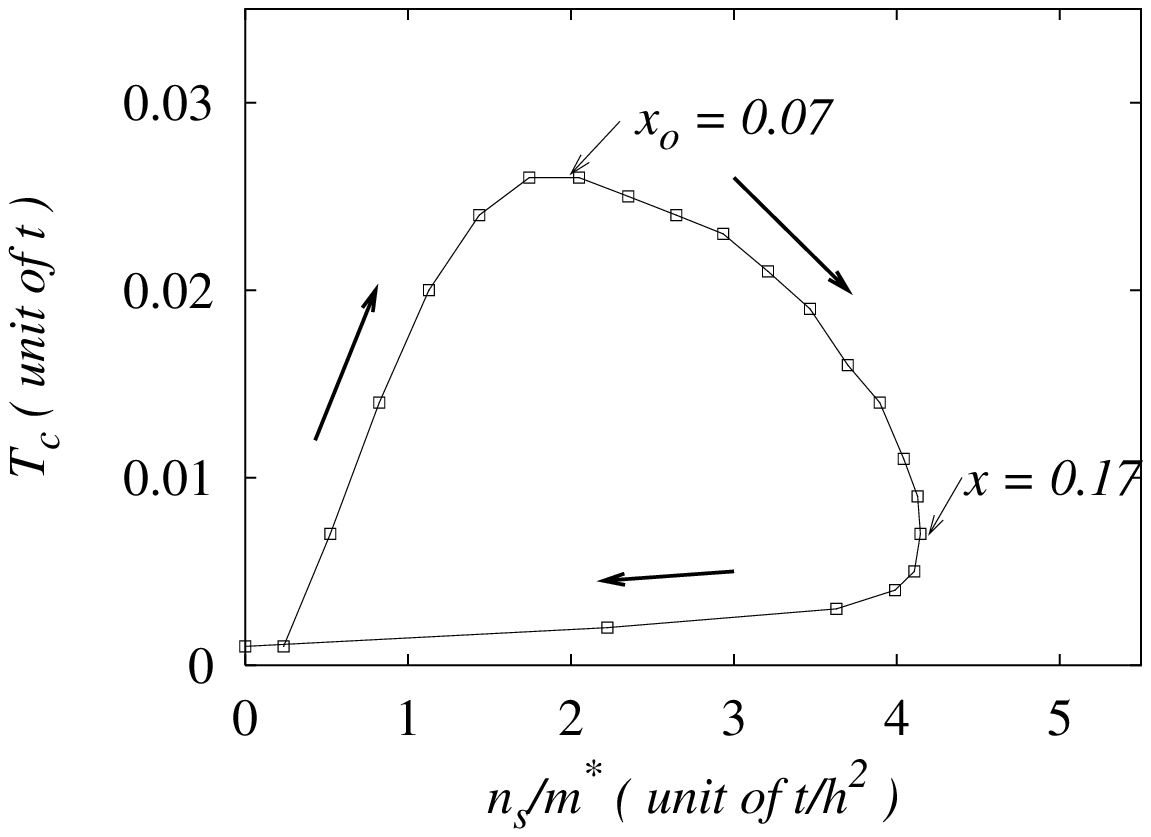}
\label{fig:6}
\caption{}
\end{figure}

\newpage
\begin{figure}
        \includegraphics{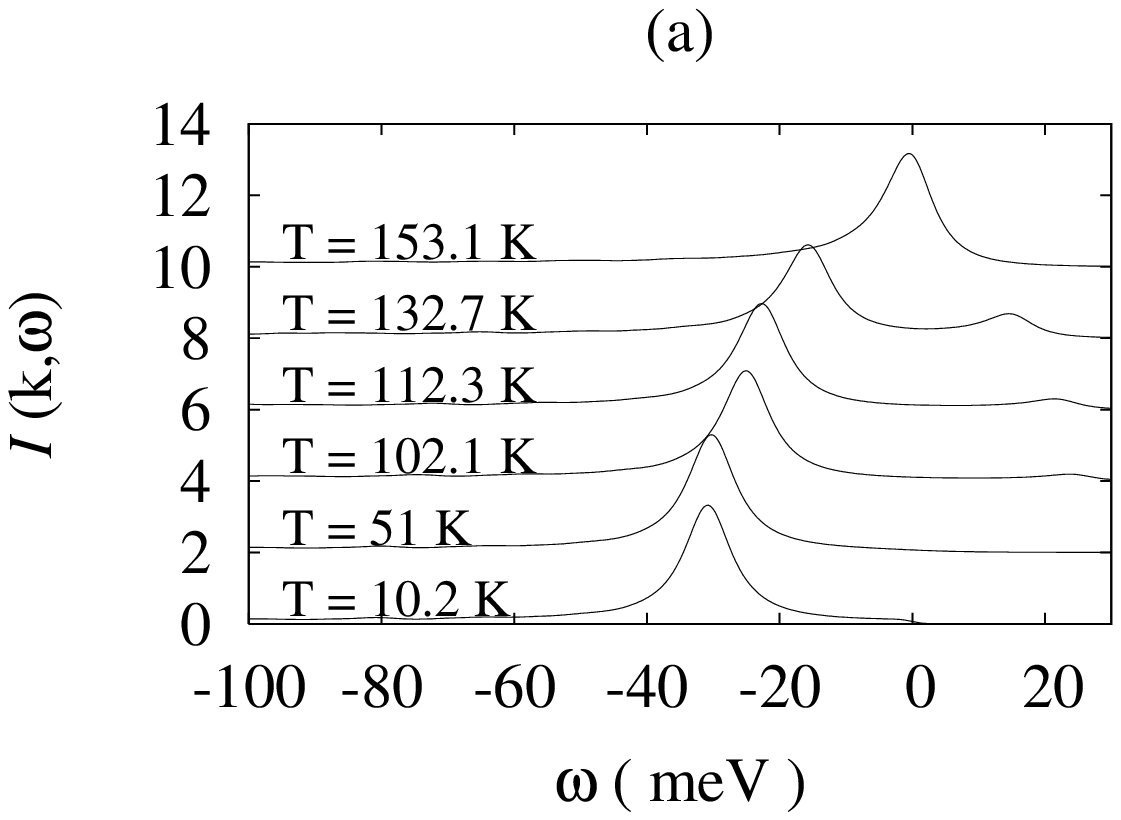}
        \includegraphics{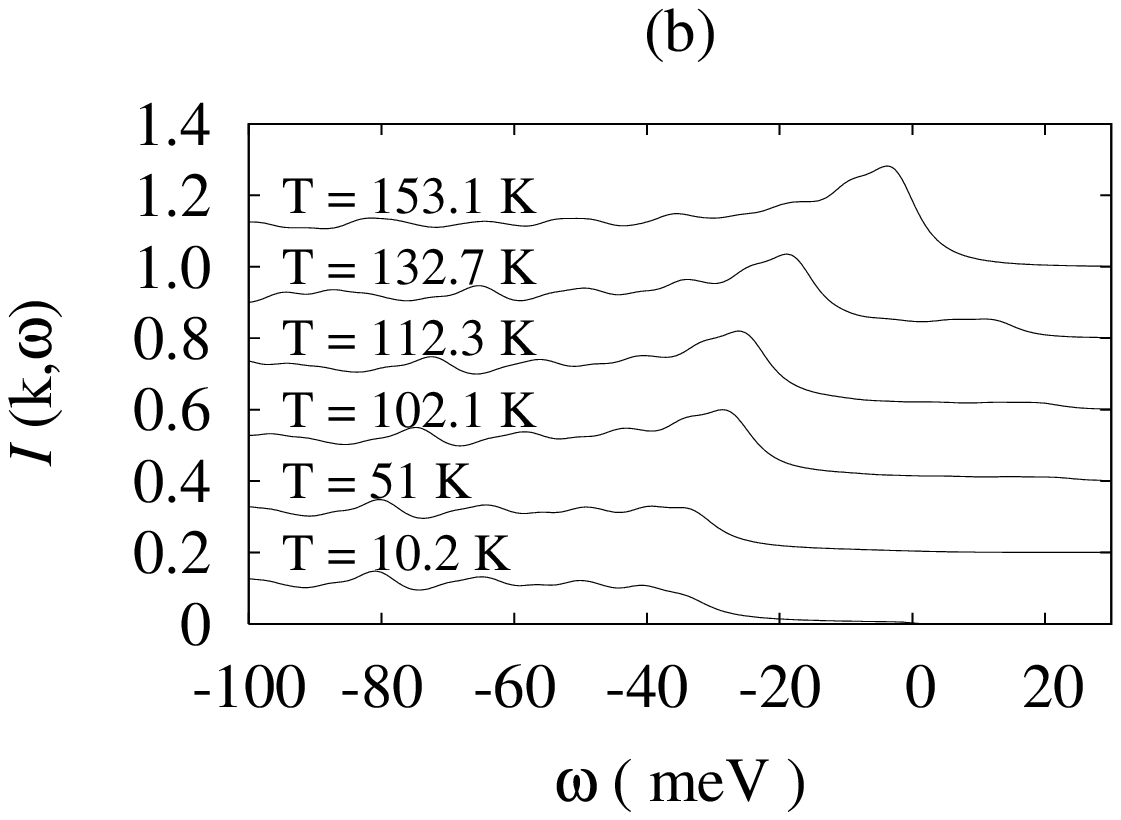}
\label{fig:7}
\caption{}
\end{figure}

\begin{figure}
        \includegraphics{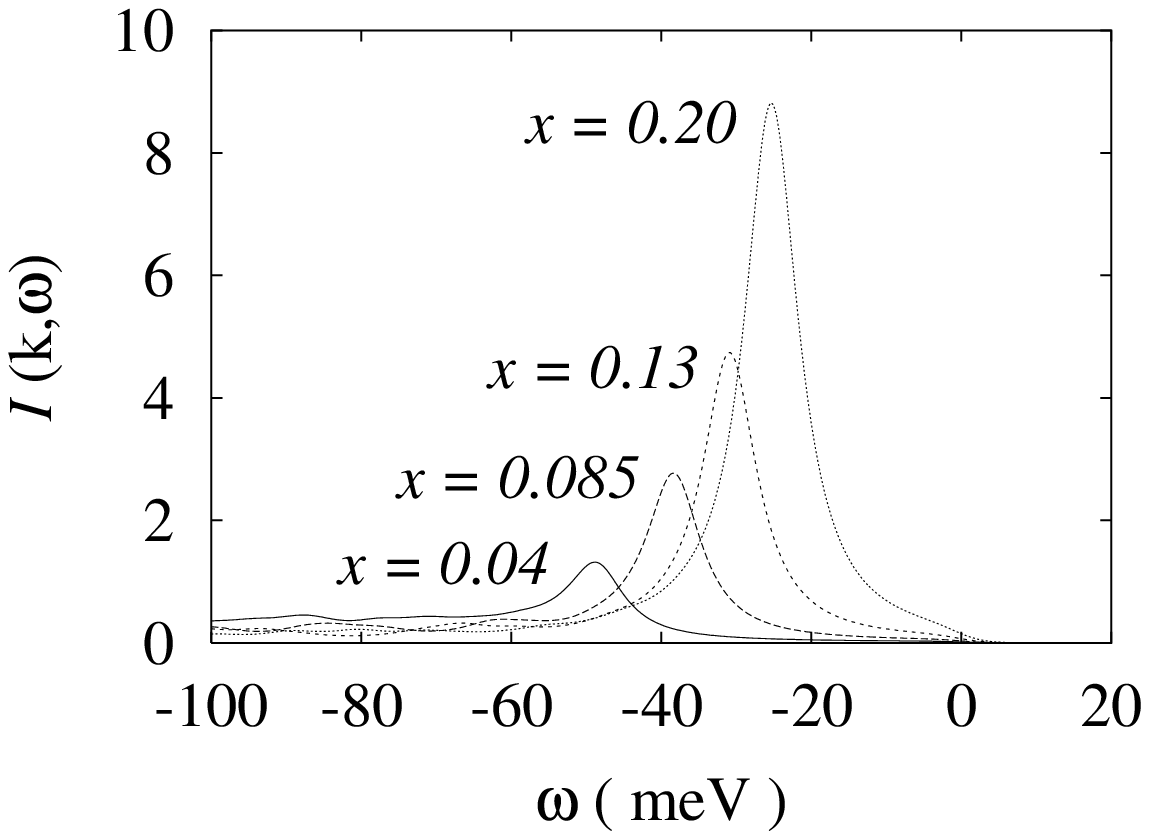}
\label{fig:8}
\caption{}
\end{figure}

\begin{figure}
        \includegraphics{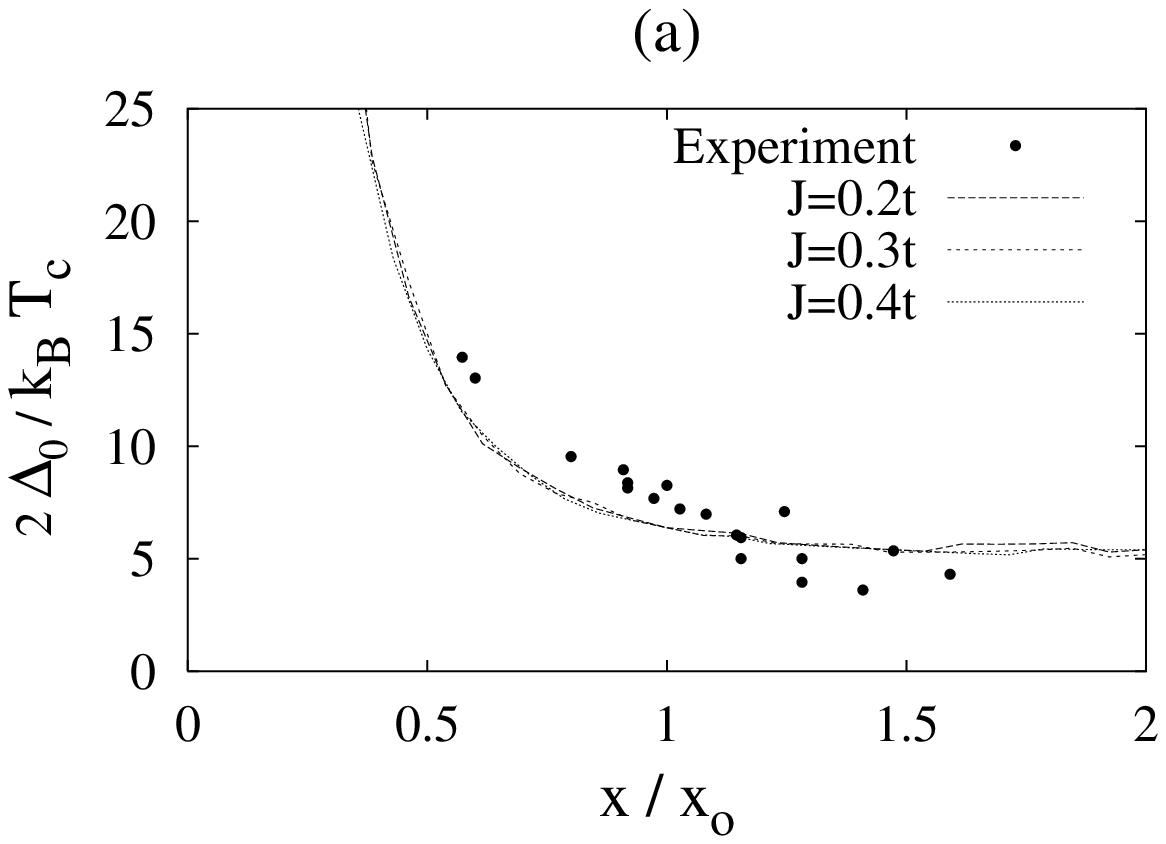}
        \includegraphics{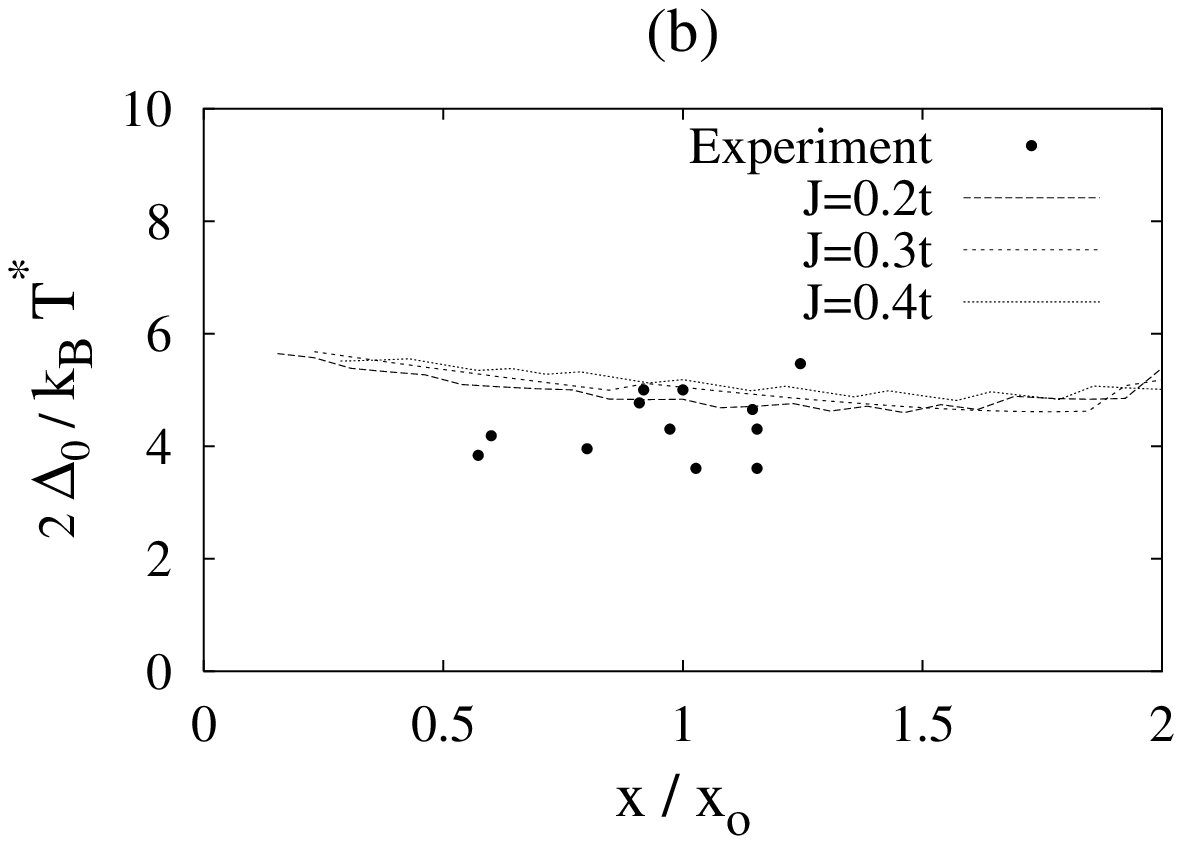}
\label{fig:9}
\caption{}
\end{figure}

\end{widetext}

\end{document}